\documentclass[aps, prd, twocolumn, lengthcheck, superscriptaddress, showpacs, letterpaper, nofootinbib]{revtex4-1}

\usepackage{amsmath,amssymb}
\usepackage{graphicx,bm}
\usepackage{slashed}
\usepackage{dcolumn}
\usepackage{epstopdf}
\usepackage{ulem} 
\usepackage[usenames]{color}
\usepackage{mathrsfs}
\def\del{\partial}
\def\be{\begin{eqnarray}}
\def\ee{\end{eqnarray}}

\allowdisplaybreaks

\begin{document}

\title{Scale-Chiral Effective Field Theory for Nuclear Interactions in the Veneziano Limit}

\author{Yan-Ling Li}

\affiliation{College of Physics, Jilin University, Changchun,
130012, China}

\author{Peng-Sheng Wen}

\affiliation{College of Physics, Jilin University, Changchun,
130012, China}

\author{Yong-Liang Ma}
\email{yongliangma@jlu.edu.cn}
\affiliation{Center for Theoretical Physics and College of Physics, Jilin University, Changchun,
130012, China}

\author{Mannque Rho}
\email{mannque.rho@cea.fr}
\affiliation{Institut de Physique Th\'eorique,
CEA Saclay, 91191 Gif-sur-Yvette c\'edex, France }

\date{\today}
\begin{abstract}
Following Golterman and Shamir~\cite{GS}, we develop scale-chiral perturbation theory in the large $N_c$ and large $N_f$ Veneziano limit that incorporates both light-quark baryons and hidden local symmetric bosons and derive a leading-order scale symmetry Lagrangian applicable in nuclear physics. {Some applications in the medium-free  space and baryonic matter are discussed.}
\end{abstract}


\maketitle


\section{introduction and some results}

\label{sec:intro}

The notion of a dilaton for the scalar field with a mass $\sim 600$ MeV, a pseudo-Nambu-Goldstone (pNG) boson of spontaneously broken scale symmetry, is found to be very portent in nuclear physics although whether or not an infrared (IR) fixed point exists in QCD for the number of flavors appropriate for nuclear  dynamics, $n_f\approx 3$,  is far from settled~\cite{MaRhoBook}. While there is growing evidence, in particular in lattice calculations, for an IR  fixed point in QCD-like gauge theories for a large number of flavors, say, $N_f=8$~\cite{Aoki:2014oha}, there are no compelling theoretical arguments either for or against the viability of the notion in QCD relevant to nuclei and nuclear matter. The scalar state light enough to be considered pNG boson is $f_0(500)$ listed in the particle data booklet.  There are no-go theorems in the literature although none of them is fatal. One of the most serious difficulties for identifying $f_0$ as a dilaton is its large width comparable to the mass, much too big to be accommodated as a Nambu-Goldstone (NG) boson~\cite{Yamawaki:2016qux}. The key question raised is how to incorporate the possible scale symmetry, inevitably explicitly broken if present, into a scheme that renders a systematic (high order) effective field theory {(EFT)} treatment.

On the other hand, there is nothing convincing that suggests  there cannot be an  IR fixed point structure in nonperturbative QCD for $N_f < 8$. In fact in holographic QCD, both infrared regime with momentum scale $Q < 1$ GeV and ultraviolet regime with $Q> 1$ GeV can have scale invariance, the former nonperturbatively and the latter perturbatively~\cite{Brodsky}. There is also a numerical stochastic perturbation  calculation that ``votes" for an IR fixed point for $N_f=2$~\cite{Horsely}. In this situation of overwhelming confusion, Crewther and Tunstall~\cite{CT} made a daring proposal for what may be called ``scale-chiral effective field theory" and identified $f_0(500)$ as the dilaton, a pNG boson of broken scale symmetry, put in the scheme commensurate with the three-flavor NG bosons. The proposal allows one to set up a systematic power counting {scale-chiral perturbation theory ($\chi$PT$_\sigma$)} generalizing the standard three-flavor chiral perturbation theory   to include the dilaton together with the octet NG bosons $\pi$. In addition to the usual power counting in chiral expansion in {three-flavor chiral perturbation theory, there is power counting associated with scale symmetry. For this one expands the beta function for the gauge coupling $\alpha_s=g^2/(4\pi)$ near the putative IR fixed point $\beta (\alpha_{IR})=0$ and counts  $|\alpha_s-\alpha_{IR}|=O(\del^2)$ as it is proportional to the mass squared of the pNG boson dilaton $f_0(500)$. One can then set up a systematic power counting in both scale symmetry and chiral symmetry treated on the same footing. In a recent paper (LMR)~\cite{Li:2016uzn},  the scale-chiral Lagrangian is written down to the next-to-leading order (NLO) in the scale-chiral counting including hidden local gauge fields and  baryon fields.  The resulting Lagrangian is denoted $bs$HLS, with $b$ standing for the baryon, $s$ for the dilaton scalar { and HLS for hidden local symmetry}.

The CT scheme as given in LMR~\cite{Li:2016uzn} has been adopted in applications to nuclear physics obtaining some interesting novel results . The basic premise in implementing the dilaton field in nuclear dynamics is that it makes sense to fluctuate around an IR fixed point, even if the location of the fixed point is unknown and  scale symmetry is ``hidden," and identify the dilaton as the scalar $\sigma$ of mass $m_\sigma \sim 600$ MeV figuring in nuclear physics, e.g.,  nuclear forces, relativistic mean-field treatments (\`a la Walecka) etc. Since scale symmetry, if present,  must  be broken both explicitly and spontaneously,  one may never be able to reach the putative IR fixed point anyway. So the aim is to probe for a possible signature of the putative symmetry emerging or ``unhidden" by strong correlations in nuclear processes. Unfortunately the Lagrangian given up to NLO is unwieldy with too many parameters.  However, as detailed in~\cite{Li:2017hqe} a potentially reliable approximation dubbed ``LOSS" -- standing for leading-order scale symmetry -- is made in such a way that the dilaton coupling to the matter fields is made, with the help of the conformal compensator field, scale-invariant  with all scale symmetry breaking put in the dilaton potential.

This LOSS approximation to the scale-chiral Lagrangian incorporating hidden local symmetric vector mesons ${\cal V}_\mu =(\rho_\mu,\omega_\mu)$ and octet baryons written in LMR~\cite{Li:2016uzn} is found to make certain predictions  for phenomena in finite nuclei and dense matter hitherto unexplained theoretically.   In applying the scale-invariant baryon hidden local symmetric Lagrangian ($bs$HLS) to nuclear dynamics,  the ``bare" parameters of $bs$HLS  are endowed with in-medium condensates of QCD at the matching scale near the chiral scale $\Lambda_\chi$, thus carrying intrinsic QCD density scaling in the Lagrangian. Applied to finite nuclei, the EFT Lagrangian is found to make several surprising predictions~\cite{Li:2017udr}. For example, it provides a simple solution to the long-standing (four-and-half-decades old) puzzle of the effective Gamow-Teller coupling constant in shell model, $g_A^\ast=1.0\pm 0.1$, it verifies  the  ``chiral filter hypothesis" made in 1970's based on soft-pion theorems, a scale-chiral symmetry derivation of Brown-Rho scaling~\cite{Li:2017hqe} etc.

Applied to dense matter relevant to compact stars~\cite{PKLR,PKLMR}, it can account reasonably well, with only a few parameters, for the bulk properties of massive $\sim 2$-solar mass compact stars.  What is however more significant is that it makes a startling novel prediction that the sound velocity of massive neutron stars approaches that of what is known as ``conformal sound velocity" $v_s^2/c^2=1/3$ at a density $\sim 3n_0$ (where $n_0$ is the normal nuclear matter density) and stays unchanged up to the maximum density present in the interior of the star, $\sim (6-7)n_0$.  This is surprising since it is known that the ``conformal sound velocity" -- normally associated with conformal invariance -- setting in and staying at non-asymptotic densities is {\it impossible} in standard nuclear physics descriptions unless non-hadronic degrees of freedom are introduced. It turns out that what represents non-hadronic degrees of freedom responsible  in this theory is the half-skyrmion phase encoded in $bs$HLS with a non-vanishing trace of energy-momentum tensor, hence with conformal invariance broken. It is the possible emergence of parity-doubling symmetry with the half-skyrmions~\cite{PKLMR} that is the cause of the sound velocity approaching the conformal value. {\it This implies that scale symmetry must emerge  in dense medium, independently of whether or not it is present in QCD in the vacuum.}

A spin-off of the approach for dense matter  is the possible impact on the tidal deformability $\Lambda$ in the gravity waves recently observed in coalescing neutron stars GW170817. The mechanism that plays a key role for the emergent conformal velocity, namely, i.e., the half-skyrmions, affects the nuclear symmetry energy in such a way that $\Lambda$ is brought down below 800~\cite{PLMR-GW}.  It is plausible that the precise pinning down of $\Lambda$ in the future measurements could provide the location of the topology change that brings in half-skyrmions in dense matter. At present,  there is no known way, experimental or theoretical,  to determine this density.

What transpires from what's observed in nuclear medium  strongly suggests that it does make sense to associate the role of the dilaton in nuclear dynamics as an emergent symmetry phenomenon, a symmetry which may not be  present in QCD proper. So the question we are interested in addressing is what makes the phenomena in nuclear dynamics ``mimic" scale symmetry spontaneously broken with a massive dilaton playing  the important role as a scalar  in nuclear dynamics?

The most illuminating suggestion for an answer to this question is the argument by Yamawaki that the Standard Model has hidden scale symmetry~\cite{Yamawaki:2016qux}. Starting with linear sigma model or equivalently Nambu-Jona-Lasinio model to which the standard Higgs model is equivalent, one can show by suitable redefinitions of chiral fields that with the  coupling constant of a four-meson interaction $\lambda$ to dial, one can arrive at nonlinear sigma model -- with no dilaton -- that captures low-energy chiral dynamics by setting setting $\lambda\to \infty$, and at a scale-invariant theory we are concerned with by setting $\lambda\to 0$.  Since HLS is gauge equivalent to nonlinear sigma model,  our LOSS approximation with $bs$HLS can be viewed as the latter with a potential that encodes scale symmetry broken, both explicitly and spontaneously. This means that it will be the dialing of $\lambda$ that governs the dynamics. It is the thesis in the application to nuclear physics that what does the dialing is the density of the matter. Then the question is: Does the exposing or unhiding of the hidden symmetry depend on the precise nature of the IR fixed point? To answer this question, as an alternative to the CT scheme on which our previous analysis is based, we analyze the alternative approach of Golterman and Shamir {(GS)}~\cite{GS}. It turns out that to the  leading order in the scale symmetry, the two approaches are identical though the physical contents may be different.

This paper is organized as follows: In sec:~\ref{sec:GS} we summarize the main features of the GS approach relevant for   this work. In sec:~\ref{sec:LOSS} we discuss the LOSS of GS and compare it with that of CT. We turn to the NLO of the GS Lagrangian in sec.~\ref{sec:bLOSS} and address some important consequences  of the scale-chiral effective theories both in the matter-free space and in baryonic medium in sec.~\ref{sec:remark}.

\section{Scale-chiral symmetry \`a la Golteman and Shamir}

\label{sec:GS}

Golterman and Shamir (GS) posit that there is an IR fixed point at a critical number of flavors $N_f^c$ and consider approaching $N_f^c$ from below. This is similar to approaching the ``sill" to the conformal window for large $N_f$ as is done for dilatonic Higgs. Here one simply assumes that there is such $N_f^c$, not necessarily as large as for dilatonic Higgs.  GS develop a scale-chiral symmetric EFT involving a dilaton and pions taking the limit of both $N_f$ and $N_c$ going to $\infty$ with $N_f/N_c$ held fixed $n_f=N_f/N_c$,  which is what is known as the Veneziano limit~\cite{Veneziano}. One can think of $N_f$ being large as $N_c$ is taken to be `large" as in large $N_c$ QCD which has been found to be quite powerful in certain cases even in nuclear physics. In what follows, we adopt this GS scheme, which is quite different physics-wise from the CT scheme, set up the systematic power counting up to NLO incorporating baryons and HLS fields as was done in LMR for the CT approach. We will then show that in the leading oder in scale-chiral symmetry, we get exactly the same LOSS Lagrangian as in the CT scheme.  This suggests that the property of LOSS in medium is generic, independent of the nature of how scale symmetry arises. This renders eminently sensible the notion of emergent scale symmetry and the role of dilaton for the scalar in dense baryonic matter.

GS assume that in the Veneziano limit, the beta function for $\alpha_s$ scales as
\be
\beta(N_c\alpha_s)=O(\Delta) +O(1/N_c),
\ee
where  $\Delta=|n_f^c-n_f|$. With the chiral counting taken the same as in {chiral perturbation theory, the power counting with scale symmetry is then given by $\Delta\sim O(\del^2)$ -- in place of $|\alpha_s-\alpha_{IR}|$ in the CT -- in expanding around the IR fixed point.  Thus the scale-chiral counting is
\be
O(p^2)\sim O(\del^2)\sim O(m_\pi^2)\sim O(\Delta).
\ee

We take the scale-invariant Lagrangian of GS~\cite{GS}\footnote{Note that we are using the notations that are different from those of GS. They are closer to the notations used in LMR~\cite{Li:2016uzn} for the CT scheme.}
\begin{eqnarray}
{\cal L}_{\chi{\rm PT}_\sigma}^{\rm LO;GS} & = & {\cal L}_{\chi{\rm PT}_\sigma;\pi}^{\rm LO; GS} + {\cal L}_{\chi{\rm PT}_\sigma;\sigma}^{\rm LO; GS} + {\cal L}_{\chi{\rm PT}_\sigma;m}^{\rm LO; GS} + {\cal L}_{\chi{\rm PT}_\sigma;d}^{\rm LO; GS}, \nonumber\\
\label{eq:GSL}
\end{eqnarray}
with
\begin{subequations}
\begin{eqnarray}
{\cal L}_{\chi{\rm PT}_\sigma;\pi}^{\rm LO;GS} & = & \frac{f_\pi^2}{4}V_\pi(\sigma - \eta) \left(\frac{\chi}{f_\sigma}\right)^2 {\rm Tr}\left( \partial_\mu U \partial^\mu U^\dagger \right),\label{eq:GSLp}\\
{\cal L}_{\chi{\rm PT}_\sigma;\sigma}^{\rm LO;GS} & = & \frac{1}{2}V_\sigma(\sigma - \eta) \partial_\mu \chi \partial^\mu \chi , \label{eq:GSt}\\
{\cal L}_{\chi{\rm PT}_\sigma;m}^{\rm LO;GS} & = &{} \frac{f_\pi^2}{4}V_M(\sigma - \eta) \left(\frac{\chi}{f_\sigma}\right)^{3-\gamma_m} {\rm Tr}\left( \mathcal{M}^\dagger U + h.c. \right), \nonumber\\
\label{eq:GSm}\\
{\cal L}_{\chi{\rm PT}_\sigma;d}^{\rm LO;GS} & = & \left(\frac{\chi}{f_\sigma}\right)^4V_d(\sigma - \eta) ,\label{eq:GSd}
\end{eqnarray}
\end{subequations}
where $U$ is the chiral (NG pion) field  $U(x) = \exp \left(2i\pi_a T_a/f_\pi\right)$  transforming  under chiral symmetry $SU_L(N_f)\times SU_R(N_f)$ as $
U(x) \to g_L U(x) g_R^\dagger$ with $g_{L,R}\in SU(N_f)_{L,R}$ for $N_f$ light flavors and  $\cal M$ is the pNG  boson mass matrix for explicit chiral symmetry breaking with $\gamma_m$ compensating the anomalous dimension of the quark mass term. Here
\begin{eqnarray}
\chi(x) & = & f_\sigma e^{\sigma/f_\sigma}
\end{eqnarray}
is the conformal compensator field of mass dimension 1 transofming linearly under scale transformation as
\begin{eqnarray}
\chi(x) & \to & \lambda \chi(\lambda^{-1}x),
\end{eqnarray}
where $\sigma$ is the dilaton field that transforms nonlinearly as
\begin{eqnarray}
\sigma(x) & \to & \sigma(\lambda^{-1}x) + f_\sigma \ln \lambda.
\end{eqnarray}
The ``invariant potentials" $V_{\pi,\sigma,M,d}$ are unknown potentials that are functionals of the dilaton field $\sigma$ and the spurion field $\eta$ that transforms like $\sigma$. These potentials,  chiral invariant by construction, are scale-invariant since the inhomogeneous terms in the transformations of $\sigma$ and $\eta$ cancel.  It turns out that the quantities dependent on $(\sigma-\eta)$ are  suppressed by the power of $\Delta$ which is a small quantity in the Veneziano limit~\cite{GS}. Now the trace anomaly, i.e., explicit breaking of scale symmetry, is generated, while chiral symmetry remains intact, when the external source $\eta$ is turned off.  To make a systematic expansion in $\Delta$, one  expands the potential $V_I$ terms of the argument $\sigma-\eta$
\begin{eqnarray}
V_I & = & \sum_{n=0}^\infty V_I^{(n)}=\sum_{n=0}^\infty\frac{c_{I,n}}{n!}(\sigma-\eta)^n ,
\end{eqnarray}
and then makes explicit the dependence on  $(n_f - n_f^\ast)^n$ by expanding $c_{I,n}$ as
\begin{eqnarray}
c_{I,n} &= & \sum_{k=0}^\infty \tilde{c}_{I,nk}(n_f - n_f^\ast)^k .
\label{eq:powerCI}
\end{eqnarray}
It was shown in \cite{GS} by matching correlation functions of the low-energy effective theory to those of the microscopic theory that $V_I^{(n)}$ scales as
\be
V_I^{(n)}(\sigma, \eta=0)=O(\Delta^n)
\ee
with $\tilde{c}_{I,nk} = 0$ for $k < n$. It should be noted that  in contrast to ordinary chiral perturbation theory in which the low-energy constants are $O(p^0)$ in the chiral counting, the coefficients of the scale-invariant potentials {$c_{I,n}$} carry power counting. This is the same as in the  Crewther and Tunstall approach~\cite{CT} with  a nonperturbative IR fixed point in QCD.

\section{Leading-order scale symmetry}

\label{sec:LOSS}

We first look at the leading-order scale-chiral symmetry in this GS scheme.

The terms multiplying $V_{\pi,\sigma,M}$ are of $O(p^2)$ in the chiral counting. Therefore to the LO in scale-chiral counting,
we can take  $V_\pi = V_\sigma = V_M = 1$ since  the correction terms are suppressed by the powers of $\Delta$ which is a small parameter in the Veneziano limit.
Thus we have
\begin{subequations}
\begin{eqnarray}
{\cal L}_{\chi{\rm PT}_\sigma;\pi}^{O(p^2)} & = & \frac{f_\pi^2}{4} \left(\frac{\chi}{f_\sigma}\right)^2 {\rm Tr}\left( \partial_\mu U \partial^\mu U^\dagger \right), \label{pi}\\
{\cal L}_{\chi{\rm PT}_\sigma;\sigma}^{O(p^2)} & = & \frac{1}{2} \partial_\mu \chi \partial^\mu \chi, \label{sigmap}\\
{\cal L}_{\chi{\rm PT}_\sigma;m}^{O(p^2)} & = &{} \frac{f_\pi^2}{4} \left(\frac{\chi}{f_\sigma}\right)^{3-\gamma_m} {\rm Tr}\left( \mathcal{M}^\dagger U + U^\dagger \mathcal{M} \right).\label{m}
\label{sigma}
\end{eqnarray}
\end{subequations}
Now as for the term (\ref{eq:GSd}) which gives the dilaton potential, scale symmetry is not broken spontaneously if one takes $V_d=1$  because the dilaton potential ${\cal V}\equiv \big(\frac{\chi}{f_\sigma}\big)^4 V_d (\sigma-\eta)\sim \big(\frac{\chi}{f_\sigma}\big)^4$ cannot pick up a non-zero vacuum expectation value. This is associated with that scale symmetry cannot be spontaneously broken unless it is explicitly broken. Thus one is required to go to $O(\Delta)\sim O(p^2)$ in scale-chiral counting,
\begin{eqnarray}
{\cal L}_{\chi{\rm PT}_\sigma;d}^{O(\Delta)} & = & \left(\frac{\chi}{f_\sigma}\right)^4\left(c_{d,0} + \tilde{c}_{d,11}(n_f - n_f^\ast)\ln \frac{\chi}{f_\sigma} \right) \label{eq:VdExp2}
\end{eqnarray}
where we have written $\sigma=\ln \chi/f_\sigma$ and, $c_{d,0}$ and $\tilde{c}_{d,11}$ are both unknown constants. In addition, $(n_f - n_f^\ast)$ which is also an unknown constant, accounts for the deviation from the IR fixed point, or equivalently, the contribution from the conformal symmetry breaking. From the saddle-point condition, the dilaton potential with lower bound is written as
\begin{eqnarray}
4c_{d,0} + \tilde{c}_{d,11}(n_f - n_f^\ast) = 0,
\end{eqnarray}
then the dilaton mass in the chiral limit is given by
\begin{eqnarray}
m_\sigma^2 & = &{} - \frac{1}{f_\sigma^2}\left[12c_{d,0} + 7\tilde{c}_{d,11}(n_f - n_f^\ast) \right]\nonumber\\
& = &{} - \frac{4}{f_\sigma^2}\tilde{c}_{d,11}(n_f - n_f^\ast) .
\end{eqnarray}
So we get the result that the coefficients $c_{d,0}$ and $\tilde{c}_{d,11}(n_f - n_f^\ast)$ are both $O(\Delta)\sim O(p^2)$ in this chiral-scale counting. Moreover using the saddle point condition, the dilaton potential with lower bound can be written as
\begin{eqnarray}
\tilde{\cal L}_{\chi{\rm PT}_\sigma;d}^{O(\Delta)} & = &{} - \frac{m_\sigma^2 f_\sigma^2}{4} \left(\frac{\chi}{f_\sigma}\right)^4\left(\ln \frac{\chi}{f_\sigma} - \frac{1}{4}\right).\label{d}
\label{eq:LdGS}
\end{eqnarray}
This is the standard Coleman-Weinberg -type dilaton potential used in the literature~\cite{Goldberger:2008zz}. Here the explicit scale-symmetry breaking sits in the $\sigma$ mass term, analogous to the pion mass term for chiral symmetry explicit breaking. The Lagrangian consisting of (\ref{pi}), (\ref{sigmap}), (\ref{m}) and (\ref{d}) corresponds to the LOSS approximation employed  in  \cite{Li:2017udr}.

It is interesting to see whether one cannot make a connection between $(n_f-n_f^\ast)$ and the anomalous dimension $\beta^\prime$ of $\rm{tr}G_{\mu\nu} G^{\mu\nu}$ that figures in the CT approach. To do this,  we rewrite the potential \eqref{eq:VdExp2}
\begin{eqnarray}
&&{\cal L}_{\chi{\rm PT}_\sigma;d}^{(\Delta)}  =  \left(\frac{\chi}{f_\sigma}\right)^4\Bigg[\left(c_{d,0} - \kappa\tilde{c}_{d,11}\right) \nonumber\\
& &\qquad\qquad\qquad\qquad\;\; {} + \kappa\tilde{c}_{d,11}\left(1 + \frac{n_f - n_f^\ast}{\kappa} \ln \frac{\chi}{f_\sigma}\right) \Bigg] \nonumber\\
 &&\simeq  \left(\frac{\chi}{f_\sigma}\right)^4\left[\left(c_{d,0} -\kappa \tilde{c}_{d,11}\right) +\kappa \tilde{c}_{d,11}\left(\frac{\chi}{f_\sigma}\right)^\frac{n_f - n_f^\ast}{\kappa} \right],
\end{eqnarray}
where $\kappa$ is a dimensionless quantity of $O(\Delta)$ such that $(n_f-n_f^\ast)/\kappa\ll 1$ is $O(1)$ in the scale-chiral counting. If one takes $(n_f-n_f^\ast)/\kappa\propto \beta^\prime$, then the GS and CT are {\it formally} the same.\footnote{For certain nonabelian gauge field theories, an exact relation between $(n_f-n_f^\ast)$ and $\beta^\prime$ must exist. This has not been worked out yet for the case we are considering since the precise nature of the IR property of GS is not known, particularly in the context with the relation to the CT theory.} This similarity can also be established by looking at the hadron coupling, such as $\sigma\pi\pi$ coupling in the chiral limit, in both approaches with $\kappa\sim m_\sigma^2/f_\sigma^2$~\cite{Li:2016uzn}. In appearance, the two approaches may be different in basic premise, but the  structure looks very similar.

\section{Going beyond the LOSS}

\label{sec:bLOSS}

\subsection{Next-to-leading order}

We have seen that to the leading order in scale symmetry, the two approaches CT~\cite{CT} and GS~\cite{GS} are related if one assumes fluctuating near the respective IR fixed point. In this section, we extend the analysis to the next-to-leading order (NLO) in chiral scale counting. We will also include both HLS and baryons. We will argue that applied to nuclear interactions, the concept of scale symmetry manifested in the process involving dilaton as the scalar relevant in nuclear physics is the same in the two schemes at the NLO.

What we will do is to detail how to go to the NLO in general in the GS scheme and then summarize what happens when HLS and baryons are involved. For the latter the argumentation made is essentially the same as what's developed in the CT scheme~\cite{Li:2016uzn}.

To the NLO, there are two terms to consider:
\begin{eqnarray}
{\cal L}_{\chi{\rm PT}_\sigma}^{\rm NLO} & = & {\cal L}_{\chi{\rm PT}_\sigma}^{{\rm GS};O(p^4)} + {\cal L}_{\chi{\rm PT}_\sigma}^{{\rm GS};{\rm LO} \times \triangle n_f},
\end{eqnarray}
where ${\cal L}_{\chi{\rm PT}_\sigma}^{O(p^4)}$ comes from counting the number of the derivative operators acting on the NGBs, here pion and $\sigma$, and the quark mass operator including its anomalous dimension $\gamma_m$ and ${\cal L}_{\chi{\rm PT}_\sigma}^{{\rm LO} \times \triangle n_f}$ comes from the NLO  of the $V_I$ in the Lagrangian \eqref{eq:GSL}.

The Lagrangian ${\cal L}_{\chi{\rm PT}_\sigma}^{{\rm GS};O(p^4)}$ can be written down straightforwardly
\begin{widetext}
\begin{eqnarray}
{\cal L}_{\chi {\rm PT}_\sigma}^{{\rm GS}; O(p^4)} & = & L_1\left[{\rm Tr}\left(\partial_\mu U^\dagger \partial^\mu U\right)\right]^2 + L_2{\rm Tr}\left(\partial_\mu U^\dagger \partial_\nu U\right){\rm Tr}\left(\partial^\mu U^\dagger \partial^\nu U\right) + L_3{\rm Tr}\left(\partial_\mu U^\dagger \partial^\mu U\partial_\nu U^\dagger \partial^\nu U\right) \nonumber\\
& &{} + L_4 \left(\frac{\chi}{f_\sigma}\right)^{1-\gamma_m}{\rm Tr}\left(\partial_\mu U^\dagger \partial^\mu U\right){\rm Tr}\left(\mathcal{M}^\dagger U + U^\dagger \mathcal{M} \right) + L_5 \left(\frac{\chi}{f_\sigma}\right)^{1-\gamma_m}{\rm Tr}\left[\partial_\mu U^\dagger \partial^\mu U\left(\mathcal{M}^\dagger U + U^\dagger \mathcal{M} \right)\right] \nonumber\\
& &{} + L_6 \left(\frac{\chi}{f_\sigma}\right)^{2(3-\gamma_m)}\left[{\rm Tr}\left(\mathcal{M}^\dagger U + U^\dagger \mathcal{M} \right)\right]^2 + L_7 \left(\frac{\chi}{f_\sigma}\right)^{2(3-\gamma_m)}\left[{\rm Tr}\left(\mathcal{M}^\dagger U - U^\dagger \mathcal{M} \right)\right]^2 \nonumber\\
& &{}  + L_8 \left(\frac{\chi}{f_\sigma}\right)^{2(3-\gamma_m)}{\rm Tr}\left(\mathcal{M}^\dagger U\mathcal{M}^\dagger U + U^\dagger \mathcal{M} U^\dagger \mathcal{M} \right) + H_2 \left(\frac{\chi}{f_\sigma}\right)^{2(3-\gamma_m)}{\rm Tr}\left(\mathcal{M}^\dagger \mathcal{M}\right) \nonumber\\
& &{} + J_1 \partial_\nu \left(\frac{\chi}{f_\sigma}\right) \partial^\nu \left(\frac{\chi}{f_\sigma}\right) {\rm Tr}\left( \partial_\mu U \partial^\mu U^\dagger \right) + J_2 \partial_\mu \left(\frac{\chi}{f_\sigma}\right) \partial^\nu \left(\frac{\chi}{f_\sigma}\right) {\rm Tr}\left( \partial_\nu U \partial^\mu U^\dagger \right) \nonumber\\
& & {} + J_3\partial_\mu \left(\frac{\chi}{f_\sigma}\right) \partial^\mu \left(\frac{\chi}{f_\sigma}\right)\partial_\nu \left(\frac{\chi}{f_\sigma}\right) \partial^\nu \left(\frac{\chi}{f_\sigma}\right) \nonumber\\
& &{} + J_4 \left(\frac{\chi}{f_\sigma}\right)^{1-\gamma_m}\partial_\mu \left(\frac{\chi}{f_\sigma}\right) \partial^\mu \left(\frac{\chi}{f_\sigma}\right) {\rm Tr}\left(\mathcal{M}^\dagger U + U^\dagger \mathcal{M} \right)\label{eq:CTsChPTNLM4}
\end{eqnarray}
\end{widetext}
where $L_i$, $H_i$ and $J_i$'s are arbitrary constants.
By comparing with what's given at the same order  in Ref.~\cite{Li:2016uzn} based on CT approach, one can see that this Lagrangian is the same as ${\cal L}_{\chi {\rm PT}_\sigma}^{O(p^4)}$ in Ref.~\cite{Li:2016uzn}.

To construct $ {\cal L}_{\chi {\rm PT}_\sigma}^{{\rm LO} \times \triangle n_f}$, let us look at $V_I,~I = (\pi,\sigma,M)$. Using the power counting given above, we can extract the $O(\Delta)$ terms of $V_I$
\begin{eqnarray}
V_I(\sigma) & = & \left(\tilde{c}_{I,01} + \tilde{c}_{I,11}\ln \frac{\chi}{f_\sigma} \right)(n_f-n_f^\ast).
\end{eqnarray}
We thus obtain the following contributions to the NLO  Lagrangian
\begin{subequations}
\begin{eqnarray}
{\cal L}_{\chi {\rm PT}_\sigma;\pi}^{\rm NLO;GS} & = & \frac{f_\pi^2}{4}\left(\tilde{c}_{\pi,01} + \tilde{c}_{\pi,11}\ln \frac{\chi}{f_\sigma} \right)(n_f-n_f^\ast) \left(\frac{\chi}{f_\sigma}\right)^2 \nonumber\\
& &{} \times {\rm Tr}\left( \partial_\mu U \partial^\mu U^\dagger \right),\\
{\cal L}_{\chi {\rm PT}_\sigma;\sigma}^{\rm NLO;GS} & = & \frac{1}{2}\left(\tilde{c}_{\sigma,01} + \tilde{c}_{\sigma,11}\ln \frac{\chi}{f_\sigma} \right)(n_f-n_f^\ast) \nonumber\\
&& \partial_\mu \chi \partial^\mu \chi ,\\
{\cal L}_{\chi {\rm PT}_\sigma;m}^{\rm NLO;GS} & = &{} \frac{f_\pi^2}{4}\left(\tilde{c}_{M,01} + \tilde{c}_{M,11}\ln \frac{\chi}{f_\sigma} \right)(n_f-n_f^\ast) \nonumber \\
& &{} \times \left(\frac{\chi}{f_\sigma}\right)^{3-\gamma_m} {\rm Tr}\left( \mathcal{M}^\dagger U + h.c. \right).
\end{eqnarray}
\end{subequations}
In this Lagrangian, the $n_f - n_f^\ast$,  $O(p^2)$ in the scale-chiral counting, is the key quantity in the GS theory in close analogy to $\Delta\alpha_s$ in the CT theory and it should be replaced with the following $O(p^2)$ chiral, parity and Lorentz invariant quantities
\begin{eqnarray}
{\rm Tr}(\partial_\mu U \partial^\mu U^\dagger), \quad \partial_\mu \chi \partial^\mu \chi, \quad {\rm Tr}(\mathcal{M}^\dagger U + h.c.).
\end{eqnarray}
Consequently the $\tilde{c}_{I,01}$ terms can be combined into the ${\cal L}_{\chi {\rm PT}_\sigma}^{{\rm GS}; O(p^4)}$ term while the $\tilde{c}_{I,11}$ terms are equivalent to  ${\cal L}_{\chi {\rm PT}_\sigma}^{{\rm LO} \times \triangle \alpha_s}$ of Ref.~\cite{Li:2016uzn} in terms of $\beta^\prime$ in the CT approach.

Finally we turn to the dilaton potential term. Up to the order $(n_f - n_f^\ast)^2$, we have
\begin{eqnarray}
{\cal L}_d^{\rm LO;GS} & = & \left[\tilde{c}_{d,00} + \left(\tilde{c}_{d,01} + \tilde{c}_{d,11}\ln \frac{\chi}{f_\sigma}\right)(n_f-n_f^\ast) \right. \nonumber\\
& & \left.\;{} + \left(\tilde{c}_{d,02} + \tilde{c}_{d,12}\ln \frac{\chi}{f_\sigma} \right.\right. \nonumber\\
& & \left.\left.\qquad{}
+ \tilde{c}_{d,22}\left(\ln \frac{\chi}{f_\sigma}\right)^2\right)(n_f-n_f^\ast)^2\right]\left(\frac{\chi}{f_\sigma}\right)^4.
\nonumber\\
\end{eqnarray}
One can see that, after the rearrangement of different terms, the dilaton potential in GS has the same form as that  in CT with $\beta^\prime$ expanded to to the second order of $\Delta \alpha_s = \alpha_s - \alpha_{\rm IR}$.

\subsection{Introducing hidden local symmetry and baryons}

Generalizing the scale-chiral Lagrangian of GS written above to hidden local symmetric bosons and baryons follows the same procedure of Ref.~\cite{Li:2016uzn}, so we won't give the details here. It suffices to summarize the results.  This we will do for the HLS fields. Given the HLS Lagrangian, putting scale symmetry is done straightforwardly using the conformal compensator field.

For completeness we define the relevant quantities~\cite{Harada:2003jx}. With the chiral field $U(x)=\xi_L^\dag(x)\xi_R(x)$ transforming under chiral transformation $\xi_{L,R}\mapsto\xi'_{L,R}=h(x)\xi_{L,R}(x) g^\dag_{L,R}$, the appropriate quantities are the parallel and perpendicular Maurer-Cartan 1-forms
$\hat\alpha_{\perp\mu}=(D_\mu\xi_{R}\cdot \xi_{R}^\dag-D_\mu\xi_{L}\cdot \xi_{L}^\dag)/2i,\ \
\hat\alpha_{\parallel\mu}=(D_\mu\xi_{R}\cdot \xi_{R}^\dag+D_\mu\xi_{L}\cdot \xi_{L}^\dag)/2i$
with the covariant derivative defined as $D_\mu=\partial_\mu-i V_\mu$.  To include the explicit chiral symmetry breaking effect, we define $\hat{\mathcal{M}}(x)\equiv\xi_L\mathcal{M}\xi^\dag_R$ which  transforms covariantly. The basic building blocks in HLS $\hat\alpha_{\perp\mu},\hat\alpha_{\parallel\mu}$ and $\frac{1}{g}V_{\mu\nu}$ have chiral order $O(p)$ while $\hat{\mathcal{M}}(x)$ has chiral order $O(p^2)$.

To the LO, the scale-symmetric HLS Lagrangian is
\begin{eqnarray}
\mathcal{ L}_{\rm{HLS}_\sigma}^{\rm LO;GS} & = & \mathcal{L}_{\rm{HLS}_\sigma;\pi}^{\rm LO;GS} + \mathcal{L}_{\rm{HLS}_\sigma;\rho}^{\rm LO;GS} + \mathcal{L}_{\rm{HLS}_\sigma;kin}^{\rm LO;GS} \nonumber\\
& &{}+ \mathcal{L}_{\rm{HLS}_\sigma;\sigma}^{\rm LO;GS} + \mathcal{L}_{\rm{HLS}_\sigma;m}^{\rm LO;GS}+\mathcal{L}_{\rm{HLS}_\sigma;d}^{\rm LO;GS}.
\label{eq:LOHLSGS}
\end{eqnarray}
The last three terms are of the same form as Eqs.~\eqref{eq:GSt}, \eqref{eq:GSm} and \eqref{eq:GSd} with the potentials $V(\sigma-\tau)$ replaced by $V_{\rm HLS} (\sigma-\tau)$. As for the first three terms, the scale invariant potentials $V_{\rm HLS}(\sigma-\tau)$s can be set equal to 1 as discussed above and one gets
\begin{subequations}
	\begin{align}
	\mathcal{L}_{\rm{HLS}_\sigma;\pi}^{O(p^2);\rm GS} = & f_\pi^2 \left(\frac{\chi}{f_\sigma}\right)^2{\rm tr}(\hat\alpha_{\perp\mu} \hat\alpha_\perp^\mu), \label{eq:LOHLSpiGS}\\
	\mathcal{L}_{\rm{HLS}_\sigma;\rho}^{O(p^2);\rm GS}
	= & af_\pi^2\left(\frac{\chi}{f_\sigma}\right)^2{\rm tr}(\hat\alpha_{\parallel\mu} \hat\alpha_\parallel^\mu), \label{eq:LOHLSrhoGS} \\
	\mathcal{L}_{\rm{HLS}_\sigma;kin}^{O(p^2);\rm GS} =&{}-\frac1{2g^2} {\rm tr}(V_{\mu\nu}V^{\mu\nu}).\label{eq:LOHLSkinGS}
	\end{align}
\end{subequations}

Now to the NLO, there are two terms as in the case of  $\chi$PT$_\sigma$,
\begin{eqnarray}
{\cal L}_{{\rm HLS}_\sigma}^{\rm NLO} & = & {\cal L}_{{\rm HLS}_\sigma}^{{\rm LO} \times \Delta n_f} + {\cal L}_{{\rm HLS}_\sigma}^{O(p^4)}.
\end{eqnarray}
For the first term as for  $\chi$PT$_\sigma$, we write the potential to $O(p^2)$
\begin{eqnarray}
V_{\rm{HLS}_\sigma;I}=(n_f-n_f^\ast)\left(\tilde{h}_{I,01}+\tilde{h}_{I,11}\ln \frac{\chi}{f_\sigma}\right).
\end{eqnarray}
Here $h$ replaces $c$ as the coefficients for HLS$_\sigma$. Then the Lagrangian ${\cal L}_{{\rm HLS}_\sigma}^{{\rm LO} \times \Delta n_f}$ can be written as
\begin{eqnarray}
{\cal L}_{{\rm HLS}_\sigma}^{{\rm LO} \times \Delta n_f} & = & {\cal L}_{{\rm HLS}_\sigma;\pi}^{{\rm LO} \times \Delta n_f} + {\cal L}_{{\rm HLS}_\sigma;\rho}^{{\rm LO} \times \Delta n_f} + {\cal L}_{{\rm HLS}_\sigma;{\rm kin}}^{{\rm LO} \times \Delta n_f}\nonumber\\
& &{} + {\cal L}_{{\rm HLS}_\sigma;\sigma}^{{\rm LO} \times \Delta n_f} + {\cal L}_{{\rm HLS}_\sigma;m}^{{\rm LO} \times \Delta n_f} + {\cal L}_{{\rm HLS}_\sigma;d}^{{\rm LO} \times \Delta n_f},
\end{eqnarray}
where the first three terms are of the same form as  \eqref{eq:GSt}, \eqref{eq:GSm} and \eqref{eq:GSd} and the next three terms are
\begin{eqnarray}
{\cal L}_{{\rm HLS}_\sigma;\pi}^{{\rm LO} \times \Delta n_f} & = & f_\pi^2 \left(\tilde h_{\pi,01}+\tilde h_{\pi,11}\ln\frac{\chi}{f_\sigma} \right)(n_f-n_f^*)\nonumber\\
& &{} \times \left(\frac{\chi}{f_\sigma}\right)^2 {\rm tr}(\hat\alpha_{\perp\mu} \hat\alpha_\perp^\mu),\nonumber \\
{\cal L}_{{\rm HLS}_\sigma;\rho}^{{\rm LO} \times \Delta n_f}
& = & a f_\pi^2 \left(\tilde h_{\kappa,01}+\tilde h_{\kappa,11}\ln\frac{\chi}{f_\sigma} \right)(n_f-n_f^*)\nonumber\\
& &{}\times  \left(\frac{\chi}{f_\sigma}\right)^2{\rm tr}(\hat\alpha_{\parallel\mu} \hat\alpha_\parallel^\mu),\nonumber\\
{\cal L}_{{\rm HLS}_\sigma;{\rm kin}}^{{\rm LO} \times \Delta n_f} & = &{} -\frac1{2g^2}\left(\tilde h_{v,01}+\tilde h_{v,11}\ln\frac{\chi}{f_\sigma} \right)(n_f-n_f^*) \nonumber\\
& &{} \times {\rm tr}(V_{\mu\nu}V^{\mu\nu}).
\end{eqnarray}
Again the quantity $(n_f-n_f^\ast)$ can be substituted by  the following Lorentz invariant $O(p^2)$ quantities
\begin{align}
&\partial_\mu\chi\partial^\mu\chi,\;\; \Box\chi,\;\; {\rm tr}(\hat{\alpha}_{\parallel \mu}\hat{\alpha}^\mu_{\parallel}) ,\;\; {\rm tr}(\hat{\alpha}_{\perp \mu}\hat{\alpha}^\mu_{\perp}) ,\;\; {\rm tr}\left(\hat{\mathcal{M}}+\hat{\mathcal{M}}^\dag\right).
\end{align}

Next, we turn to the ${\cal L}_{{\rm HLS}_\sigma}^{O(p^4)}$ sector. This sector includes two parts: The part which has the same form as the standard HLS in the chiral limit (since this part is already scale-invariant) and the other part arising with the dilaton field. The former part is obvious. For the latter, it can be written as
\begin{widetext}
\begin{eqnarray}
{\cal L}_{{\rm HLS}_\sigma}^{O(p^4)} & = & \left[D_{\pi,1}\left(\partial_\mu \frac{\chi}{f_\sigma}\right)^2 + D_{\pi,2}\Box\left(\frac{\chi}{f_\sigma}\right)\right]{\rm tr}(\hat\alpha_{\perp\nu} \hat\alpha_\perp^\nu) + D_{\pi,3}\partial_\mu \left(\frac{\chi}{f_\sigma}\right) \partial_\nu \left(\frac{\chi}{f_\sigma}\right){\rm tr}(\hat\alpha_\perp^\mu \hat\alpha_\perp^\nu)\nonumber\\
& &{} + \left[D_{\kappa,1}\left(\partial_\mu \frac{\chi}{f_\sigma}\right)^2+D_{\kappa,2}\Box\left(\frac{\chi}{f_\sigma}\right)\right]{\rm tr}(\hat\alpha_{\parallel\nu} \hat\alpha_\parallel^\nu) + D_{\kappa,3}\partial_\mu \left(\frac{\chi}{f_\sigma}\right) \partial_\nu \left(\frac{\chi}{f_\sigma}\right){\rm tr}(\hat\alpha_\parallel^\mu \hat\alpha_\parallel^\nu)\nonumber \\
& &{} + D_{m,1} \left(\frac{\chi}{f_\sigma}\right)^{(y-2)} {\rm tr}\left[\hat\alpha_{\perp\mu} \hat\alpha_\perp^\mu(\hat{\mathcal{M}} + \hat{\mathcal{M}}^\dag)\right] + D_{m,2} \left(\frac{\chi}{f_\sigma}\right)^{(y-2)}{\rm tr}\left[\hat\alpha_{\parallel\mu} \hat\alpha_\parallel^\mu(\hat{\mathcal{M}} + \hat{\mathcal{M}}^\dag)\right]\nonumber \\
& &{} + D_{m,3}\left(\frac{\chi}{f_\sigma}\right)^{(y-2)}{\rm tr}\left[\left(\hat\alpha_\parallel^\mu\hat\alpha_{\perp\mu}-\hat\alpha_{\perp\mu}\hat\alpha_\parallel^\mu\right)\left(\hat{\mathcal{M}} - \hat{\mathcal{M}}^\dag\right)\right], 
\end{eqnarray}
\end{widetext}
where $D_{I,n} (I=\pi, \kappa, m$ and $n=1, 2, 3)$ are the low-energy constants. Note that there are some terms independent of $\hat\alpha_{\parallel\mu}$ and $\hat\alpha_{\perp\mu}$. They have the same forms as that in $\chi$PT$_\sigma$, so it is straightforward to write down their explicit forms.

Finally, we introduce baryon. As Ref.~\cite{Li:2016uzn}, we introduce the baryon octet field $B$ which transforms under HLS as $B(x) \to h(x)B(x)h^\dagger(x)$ and use the canonical dimension of
the baryon field, i.e., under scale transformation, it transforms
as $ B(x) = \lambda^{3/2}B(\lambda^{-1}x)$ .

To the LO, that is, $O(p)$ or $O(\Delta^{1/2})$, the scale invariant potential $V_{bs{\rm HLS}}(\tau - \sigma)$s can also be set to $1$ as before. Thus the Lagrangian for baryons part takes the form:
\begin{eqnarray}
\mathcal{ L}_{bs\rm{HLS}}^{\rm LO;GS} & = & {\rm tr}({\bar B} i\gamma_\mu D^\mu B) - {\mathring{m}}_{B}\frac{\chi}{f_\sigma} {\rm tr}({\bar B}B)\nonumber\\
& &{} - g_{A_1}{\rm tr}\left({\bar B}\gamma_\mu \gamma_5\{ \hat\alpha_{\perp}^{\mu},B\}\right)\nonumber\\
& &{} - g_{A_2}{\rm tr}\left({\bar B}\gamma_\mu \gamma_5[ \hat\alpha_{\perp}^{\mu},B]\right)\nonumber\\
& &{} - g_{V_1}{\rm tr}\left({\bar B}\gamma_\mu \{ \hat\alpha_{\parallel}^{\mu},B\}\right)\nonumber\\
& &{} - g_{V_2}{\rm tr}\left({\bar B}\gamma_\mu [ \hat\alpha_{\parallel}^{\mu},B]\right),
\label{eq:LLOSSB}
\end{eqnarray}
which is the LOSS of the baryonic part. In \eqref{eq:LLOSSB}, ${\mathring{m}}_{B}$ is the baryon mass in the chiral limit.

As for the NLO Lagrangian, i.e., $O(p^2)$ or $O(\Delta)$,  as in the cases of  the $\chi$PT$_\sigma$ and HLS$_\sigma$, there are two quantities contributing to the power counting: The first $\mathcal{ L}_{bs\rm{HLS}}^{{\rm LO}\times \Delta n_f;{\rm GS}}$ is from the NLO  expansion of the scale invariant potentials and the second $\mathcal{ L}_{bs\rm{HLS}}^{O(p^2)}$ is from the expansion of derivative operators and quark mass matrix.

For $\mathcal{ L}_{bs\rm{HLS}}^{{\rm LO}\times \Delta n_f;{\rm GS}}$,  only the first two terms need to be multiplied by the potentials $V_{1}(\sigma - \tau)$ and $V_{{\mathring{m}}}(\sigma - \tau)$, respectively. Therefore, we have
\begin{eqnarray}
\mathcal{ L}_{bs\rm{HLS}}^{{\rm LO}\times \Delta n_f} & = & \left(\tilde{ c}_{1,01}+\tilde c_{1,11}\ln\frac{\chi}{f_\sigma}\right)(n_f-n_f^\ast){\rm tr}({\bar B} i\gamma_\mu D^\mu B) \nonumber\\
& &{} - \left(\tilde{ c}_{{\mathring{m}},01}+\tilde c_{{\mathring{m}},11}\ln\frac{\chi}{f_\sigma}\right)(n_f-n_f^\ast){\mathring{m}}_B{\rm tr}({\bar B} B).
\nonumber\\
\end{eqnarray}
In general, this Lagrangian contributes to the $O(p^2)$ of the $bs$HLS. However, if we set
 $\tilde{ c}_{1,01} = \tilde{ c}_{{\mathring{m}},01}$ and $\tilde{ c}_{1,11} = \tilde{ c}_{{\mathring{m}},11}$, this is of the form
\begin{eqnarray}
\mathcal{ L}_{bs\rm{HLS}}^{{\rm LO}\times \Delta n_f} & \propto & (n_f-n_f^\ast){\rm tr}\left[{\bar B} \left(i\gamma_\mu D^\mu - {\mathring{m}}_B \right) B\right] \sim O(p^3).
\nonumber\\
\end{eqnarray}
{Therefore, $\mathcal{ L}_{bs\rm{HLS}}^{{\rm LO}\times \Delta n_f;{\rm GS}}$ does not contribute to the NLO $bs$HLS.}

As for $\mathcal{ L}_{bs\rm{HLS}}^{O(p^2)}$, one can readily write down the most general Lagrangian following the standard method.  

\section{Remarks}

\label{sec:remark}

Although the IR structure is presumably different in the way the putative fixed point is different, the GS scheme and CT scheme are of the same form to NLO once $\beta^\prime$ in CT and $\Delta n_f$ in GS are related. Although one can write them down explicitly, both in their present form are essentially powerless in confronting Nature since there are too numerous  constants that can be determined neither phenomenologically nor theoretically. However to the LOSS  and with certain conditions imposed, they are identical and can be used to calculate both normal nuclear  and compact-star properties.

Here we illustrate two cases where one can make quantitative analyses, one in the matter-free space and, another in baryonic matter.

At the LOSS order, \eqref{eq:LOHLSpiGS} gives the $\sigma\pi\pi$ coupling of the derivative form $\sim\sigma \partial_\mu \bm{\pi} \cdot \partial^\mu \bm{\pi}$. For the on-shell dilaton, one can write the $\sigma\pi\pi$ coupling as
{
\begin{eqnarray}
{\cal L}_{\sigma\pi\pi} & = &{} \frac{1}{2}g_{\sigma\pi\pi} \sigma \bm{\pi}\cdot \bm{\pi}
\end{eqnarray}
}
with
\begin{eqnarray}
g_{\sigma\pi\pi} & = &{} \frac{m_\sigma^2}{ f_\sigma} - 2 \frac{m_\pi^2}{ f_\sigma} - \frac{(3-\gamma_m)m_\pi^2}{f_\sigma}
\label{eq:gSPP}
\end{eqnarray}
where the first two terms come from the derivative coupling between $\sigma$ and pion and the third term comes from the explicit chiral symmetry breaking term of the effective Lagrangian. The $\sigma \to \pi \pi$ width is
{
\begin{eqnarray}
\Gamma(\sigma \to \pi\pi) & = &{} \frac{3}{2} \times \frac{g_{\sigma\pi\pi}^2}{8 \pi m_\sigma^2} |P_\pi|
\end{eqnarray}
}
where $|P_\pi| = \sqrt{m_\sigma^2(m_\sigma^2 - 4 m_\pi^2)}/2 m_\sigma$ is the three momentum of the decay products. If one ignored the derivative couplings and take only the third term, one would get
{
\begin{eqnarray}
\Gamma(\sigma \to \pi\pi) & \simeq  &{} 7.6 ~{\rm MeV}
\end{eqnarray}
with $m_\sigma \simeq 500~{\rm MeV}, m_\pi \simeq 140~{\rm MeV}, f_\sigma \simeq 100~{\rm MeV}$ and $\gamma_m \simeq1$, which  is too small compared with the observed value $400$-$700$~MeV~\cite{PDG}. } However Eq.~\eqref{eq:gSPP} with the derivative coupling gives
{
\begin{eqnarray}
\Gamma(\sigma \to \pi\pi) & \simeq  &{} 145~{\rm MeV}.
\end{eqnarray}
Although this result is just about one third of the empirical value, they are of the same order. Given the large phase available,  a width fully compatible with the observation could be understood by considering loop corrections involving the $\beta^\prime$ contribution -- which plays a crucial role for the dilaton mass -- as suggested in the case of CT  and the NLO corrections from $V_\pi(\sigma-\eta)$ in the case of GS.}

A lot more interesting case is in nuclear and dense baryonic matter.

With baryons incorporated into HLS$_\sigma$, the resulting $bs$HLS Lagrangian at the LOSS approximation is found to give a fairly satisfactory account of both normal nuclear matter and dense compact-star matter. This is described in \cite{PKLR,PKLMR,PLMR-GW}.

A more striking prediction is found with the famous phenomenon of quenched $g_A$ in nuclear medium. This is described in \cite{Li:2017udr}.  We briefly describe it here as a case where the two approaches GS and CT at the LOSS are identical.

The issue is the effective Gamow-Teller (GT) coupling constant $g_A^{\rm eff}$ in nuclear matter mentioned in Section I. In the matter-free space, from precision neutron beta decay, the axial coupling constant in $g_A$ in the weak current is currently given as
\be
g_A=1.2755(11).
\ee
But in nuclei described in shell model it comes out to be
\be
g_A^{\rm eff}=1.0\pm 0.1.
\ee
This is the quenched $g_A$ problem that has challenged nuclear theorists for more than four decades.  In \cite{Li:2017udr}, it has been shown that this can be {\it very simply} explained by scale-chiral EFT in the LOSS approximation.

Since the problem could be possibly exposing a basic difference in the way scale symmetry ``emerges" in  nuclear medium  between the CT and GS, we address the problem using the CT approach. The solution described here will be at the LOSS, at which both give the same result.  So the question arises at what level of approximation the GS could differ from the CT in precisely measured quantities in nuclei, i.e., the GT transitions in nuclei?

In the CT approach~\cite{CT}, the single-particle Gamow-Teller operator in nuclear medium at small momentum transfer is given by
\be
\left[c+ (1-c)\left(\frac{f_\sigma^*}{f_\sigma}\right)^{\beta^\prime}\right] g_A\tau^{\pm}\mathbf{\sigma}
\ee
where $c$ is an unknown constant. The LOSS is obtained for $c=1$ or $\beta^\prime\to 0$ and the operator goes to
\be
g_A\tau^{\pm}\mathbf{\sigma}.
\ee
In GS however this follows immediately without any alternatives because the potential $V_I (\sigma)=1+O(\Delta)$ and the GT operator is at the LOSS given by 1. It may be that in Nature $\beta^\prime\neq 0$ and $c\neq 0$. Then there will be a difference as to how LOSS is approached in the two approaches.  Furthermore  in chiral perturbation theory in medium, two-body exchange-current contributions come at $O(p^2)$ relative to the single-particle operator and the chiral filter argument suggests that it is strongly suppressed.  Using the mean-field treatment of the LOSS Lagrangian as an approximation to Landau Fermi liquid theory, it has been shown that the Gamow-Teller transition in terms of quasiparticles at the Fermi-liquid fixed point which is equivalent to a  simple shell-model transition within single-shell model space requires 21 \% reduction in the coupling constant. This gives
\be
g_A^\ast=0.79 \times 1.2755=1.01
\ee
with practically no density dependence in the vicinity of nuclear matter density $n_0=0.16$ fm$^{-3}$. Applied to finite nuclei, this just means that the missing 21\% strength is in  nuclear correlated states strongly coupled to the initial state by the nuclear tensor forces. This result when accurately determined could give information on $c$ and $\beta^\prime$ in the scale-chiral EFT in the form of emergent symmetry.

\subsection*{Acknowlegments}

Y.~L. Ma was supported in part by National Science Foundation of China (NSFC) under Grant No. 11475071, 11747308 and the Seeds Funding of Jilin University.

\end{document}